\documentclass[]{spie}  

\newcommand{\ket}[1]{|#1\rangle} 

\usepackage{amsmath,amsfonts,amssymb}
\usepackage{graphicx}
\usepackage[colorlinks=true, allcolors=blue]{hyperref}
\usepackage{graphicx}

\newcommand{\numberlist}[2][0.8\linewidth]{%
  [\parbox[t]{#1}{\printcommalist{#2}}%
}
\newcommand{\printcommalist}[1]{%
  \begingroup\lccode`~=`,\lowercase{\endgroup\def~}{\mathcomma\penalty0 }%
  \mathcode`,="8000
  \thinmuskip=6mu plus 6mu minus 2mu
  $#1]$
}
\mathchardef\mathcomma=\mathcode`,

\title{A Programmable True Random Number Generator Using Commercial Quantum Computers}

\author[a]{Aviraj Sinha}
\author[a]{Elena R. Henderson}
\author[a]{Jessie M. Henderson}
\author[a]{Eric C. Larson}
\author[a]{\hspace{3em} Mitchell A. Thornton}
\affil[a]{Darwin Deason Institute for Cyber Security, Southern Methodist University, \hspace{7em} 6425 Boaz Lane Dallas, TX 75205, USA}

\authorinfo{Further author information: (Send correspondence to AS.)\\AS: avirajs@smu.edu, ERH: erhenderson@smu.edu, JMH: hendersonj@smu.edu, ECL: eclarson@smu.edu, MAT: mitch@smu.edu}

\begin{document} 
\maketitle

\begin{abstract}
Random number generators (RNG) are essential elements in many cryptographic systems.
True random number generators (TRNG) rely upon sources of randomness from natural processes such as those arising from quantum mechanics phenomena.
We demonstrate that a quantum computer can serve as a high-quality, weakly random source for a generalized user-defined probability mass function (PMF).
Specifically, QC measurement implements the process of variate sampling according to a user-specified PMF resulting in a word comprised of electronic bits that can then be processed by an extractor function to address inaccuracies due to non-ideal quantum gate operations and other system biases.
We introduce an automated and flexible method for implementing a TRNG as a programmed quantum circuit that executes on commercially-available, gate-model quantum computers.
The user specifies the desired word size as the number of qubits and a definition of the desired PMF. Based upon the user specification of the PMF, our compilation tool automatically synthesizes the desired TRNG as a structural OpenQASM file containing native gate operations that are optimized to reduce the circuit's quantum depth.
The resulting TRNG provides multiple bits of randomness for each execution/measurement cycle; thus, the number of random bits produced in each execution is limited only by the size of the QC.
We provide experimental results to illustrate the viability of this approach.
\end{abstract}


\section{Introduction}
\label{intro}
Many cryptographic schemes for communications, networking applications, and encryption/decryption of data rely upon the use of random number generators (RNG).
Examples include the generation of temporary passwords, encryption keys, and cryptographic seed, salt, and nonce values.
A high-quality RNG is secure and capable of producing random values at a very high rate of speed.
Secure RNGs are structured with the goal of restricting an adversary from predicting a produced value with a probability greater than chance.
Some practical applications that depend upon high-quality RNGs include Open SSL and the generation of initial sequence numbers for TCP/IP.
The latter networking applications are also examples that require an RNG to produce a digit stream at a high rate of speed.

There are two major classes of RNG: pseudo-random number generators (PRNG) and true random number generators (TRNG).
These are also referred to as deterministic random bit generators (DRBG) and non-deterministic random bit generators (NRBG) respectively~\cite{BK+:07,TB+:18}.
PRNG/DBRG are useful when an application simultaneously requires a bitstream with short-term statistics identical to a truly random stream and reproducibility of a shared secret.
Two examples are the use of a linear feedback shift register (LFSR) ~\cite{golub2013matrix} or a Mersenne twister~\cite{matsumoto1998mersenne}.
One application of PRNGs is that of a scrambling code, such as those used in the IEEE 802.11 WiFi standards in which the transmitter generates the scrambling sequence, the receiver generates the same sequence for descrambling, and a channel eavesdropper cannot distinguish the transmitted and scrambled bit sequence from a truly random bit sequence.

A TRNG/NRBG is used for applications where the produced bit stream should be secure, meaning that it cannot be predicted in advance with a chance greater than a random guess.
PRNGs are less secure than TRNGs since they have deterministic long-range correlations that may result in information leakage over time.
For this reason, most PRNGs often use TRNGs as the seed input to start the number collection.
Programmable random number generators that are cryptographically secure (CSPRNGs) must be unpredictable and backwards-secure, meaning it must be impossible to repeatedly generate a sequence given the same generator.
In theory, only a TRNG can fully meet these requirements.

\begin{figure} [ht]
\begin{center}
\begin{tabular}{c}
\includegraphics[width=0.9\textwidth]{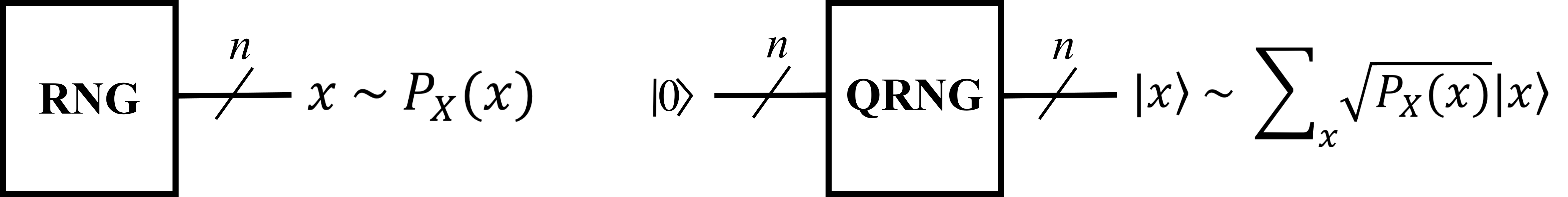}
\end{tabular}
\end{center}
\caption[example] 
{\label{fig:TRNG_structure} 
Classical and Quantum Architecture of a TRNG.}
\end{figure}

TRNGs are often based upon a natural source of randomness that is quantum mechanical in nature, resulting in a quantum random number generator (QRNG).  
Such sources include radioactive decay, semiconductor device noise, photodetector dark counts, and photon paths through semi-transparent mirrors.
High-quality TRNGs are typically in the form of application-specific devices with a fixed PMF and output bit rate.
However, due to the inherently probabilistic nature of quantum computers, it is possible to consider a TRNG in the form of a quantum circuit that executes on a programmable quantum computer.
Applying quantum computers as TRNGs for fixed PMFs is not a new idea; for example, a qubit initialized to a fiduciary state and evolved through an appropriate single qubit gate such as a Hadamard or Square-root of NOT gate serves as a single-bit TRNG with a Bernoulli PMF.
However, the ability to automatically augment this process by preparing a state generation circuit representing a PMF for a multi-qubit quantum computer allows for conveniently implementing a flexible, high-quality QRNG.
Furthermore, using multiple qubits allows the programmable QRNG to generate multiple bits per execution or, alternatively, to generate random digits with a programmable degree of resolution.
This general structure of a QRNG and classical RNG is illustrated by the block diagram in Figure~\ref{fig:TRNG_structure} for side-by-side comparison.

This work introduces a method for automatically synthesizing a programmable QRNG into an optimized and low-level quantum computing language, OpenQASM~\cite{CB+:17,CB+:20}.
Designing and implementing a state-generation circuit to represent a desired PMF is generally a complex quantum programming task for arbitrary PMFs.
However, we have developed an automated procedure wherein the desired PMF is specified in tabular form as a function using conventional fixed-point values.
This table is parsed, synthesized, and optimized by our automatic quantum design tool, \textit{MustangQ}, to produce the required state generation circuit automatically.
Therefore, the only input requirements are the number of qubits to be used and a description of the desired PMF.
This tabular form allows for any arbitrary PMF to be specified, including non-parametric PMFs.

There are many potential applications of our flexible QRNG, hereafter referred to as ``Flex-QRNG."
First, it could be used in cryptographic processes, either as a primitive element in existing and emerging cryptographic processes---including as support for emerging post-quantum cryptographic (PQC) methods--or in new, truly-quantum-based cryptographic methods.
For example, some candidate PQC codes require many random values from a binomial distribution.  It should be noted that many of the RNG applications previously mentioned are designed with the goal of producing truly random, or equally-likely, output values.  In these cases, the PMF of interest is the discrete uniform probability mass function.  While Flex-QRNG can certainly produce a quantum circuit that adheres to the uniform PMF, it can also produce circuits that adhere to other parametric, or even non-parametric, PMFs.

Second, Flex-QRNGs could be used to model an observed non-parametric distribution for simulations such as Monte Carlo methods.
Monte Carlo methods require accessible and high-speed RNGs and can speed up many complex problems by averaging random instances.
Non-log concave distributions are a subclass of nonparametric distributions that are known to be hard to sample from classically~\cite{grover2002creating}, so QRNGs can be used instead to generate distributions that are non-log concave.

Third, Flex-QRNGs provide a quantum distribution that can be processed uniquely on a quantum computer.
For example, Flex-QRNGs can be used as non-uniform priors for Grover search oracles~\cite{Gro:96}.
Grover's search algorithm can be enhanced by allowing a user to weigh some regions of the search space more than others, given that a solution is more likely to be found in that region.
Such a distribution can be referred to as a prior distribution or the estimated distribution of the search result before any calculation.
Another example is the creation of uniquely quantum processing algorithms that transform an initial state into another complex state in the Hilbert space before performing techniques such as quantum interference for any probability calculation.
Yet more examples of Flex-QRNGs are found in well-known algorithms such as Quantum Amplitude Estimation (QAE) and the Harrow-Hassidim-Lloyd (HHL) algorithm for solving linear systems.

The remainder of this paper proceeds as follows.
Sec.~\ref{sec:background} provides background on synthesizing QRNGs for arbitrary PMFs.
Then, Sec.~\ref{sec:implementation_optimization_flex_qrng_circuits} describes the theory and architecture of the Flex-QRNG circuit and the automatic synthesis tool used to produce it.
The remaining Sections demonstrate the use of the tool by generating and analyzing several Flex-QRNG circuits.
In addition to presenting circuit metrics, such as quantum gate count and quantum depth, we demonstrate circuit functionality by executing the Flex-QRNG circuits on the IBM-Q machine simulators.
Specifically, we simulate the circuits to generate non-uniform random numbers and then apply inference testing to ensure the circuit represents the defined PMF.
We conclude by briefly considering areas for future research.

\section{Background and Previous Research}\label{sec:background}
\subsection{Uniform QRNGs and Extractors}
Leveraging quantum computing hardware for generating uniform random numbers is a fairly well-developed area of research~\cite{herrero2017quantum}.
For example, the research in Refs.~\citenum{tamura2020quantum, salehi2022hybrid} has considered straightforward techniques for generating uniform hardware-independent QRNGs on quantum computers by applying Hadamard and controlled-Hadamard gates.
These past research projects also analyze the quality of the generated random numbers by applying Samuelson and Von Neumann extractors, as well as uniformity testing using simple frequency analysis for assessing the efficacy of the extraction.
Other work has explored improving the randomness of the generated numbers by using entanglement generators to group qubits as random values~\cite{jacak2020quantum}.

Additionally, a significant amount of past work has considered extractors for RNGs.
Even in the theoretically-perfect case in which the natural source of randomness has no deterministic components, the presence of the measuring or observation device will generally introduce some degree of determinism or bias.
Furthermore, the native or inherent probability mass function (PMF) arising from the quantum circuit will likely need to be transformed into a different user-desired PMF.
For this reason, the output of the weakly random source (WRS) is processed by an ``extractor," which is a mathematical function that serves two main purposes: first, it removes the deterministic component of the WRS, and second, it transforms the native PMF of the random component of the WRS into that desired by the user.
Theoretically, the extractor can be considered as a mathematical function that performs a random variable (RV) transformation.
The National Institute of Standards and Technology (NIST) has created guidelines that specify whether a candidate RNG is, in fact, cryptographically secure based upon the error rate\cite{barker2007recommendation}.
Extraction methods can transform imperfect TRNGs into NIST-acceptable TRNGs.
A great deal of work has considered extractors for both uniform and non-uniform random-number-generating circuits.
For example, Ref.~\citenum{ma2013postprocessing} discusses random-number-quality testing using min-entropy analysis and advanced extractors, such as the Trevison extractor and Toeplitz hashing.

Extraction for non-uniform distributions is non-trivial and is a subject worthy of a different paper, as many factors must be considered to meet all the test requirements of the NIST standard. Most of the past work and extractors mentioned above are designed with the goal of transforming the random component of the WRS into a uniformly distributed RV.  In contrast, the extraction method to be employed with Flex-QRNG should preserve the RV PMF as defined by the user in the input data and simply remove perturbations introduced by quantum gate systemic errors and imperfect quantum measurement devices. In addition, non-uniform PMF extractors must remove correlations that exist over large instances of time as properties of the host quantum computer change.
Therefore, we will not further discuss non-uniform PMF extraction techniques here, instead we focus on characterizing the WRS that consists of the host quantum computer and the Flex-QRNG automatically synthesized circuit.  To evaluate the WRS, we statistically test the shape of our sampled distribution by comparing the variates measured from the implemented WRS on an IBM quantum computer with the ideal PMF as specified by the input file to the Flex-QRNG synthesis tool.

\subsection{Amplitude Encoding Method for Generating Arbitrary PMFs on Quantum Computers}\label{sec:amplitude_encoding_method}

Our approach for generating Flex-QRNG circuits uses an amplitude encoding circuit structure to encode probabilities of a discrete PMF as amplitudes in the quantum states. By using the amplitudes to represent bins of a PMF distribution rather than individual data points, we can program the circuit to represent arbitrary non-parametric distributions.

As we will apply a general amplitude encoding method ~\cite{grover2002creating} to create our state representing a PMF, we summarize that procedure here.
We can use the ratio of the left and right volumes of the PMF as a parameter of the additional qubit rotation: $f(i)=\frac{\sum_{x_L^i}^{\frac{x_R^i-x_L^i}{2}} P(x) }{\sum_{x_L^i}^{x_R^i} P(x)}$.
We can then apply a rotation to each individual qubit state that splits our $|0\rangle$ and $|1\rangle$ probabilities.
The rotation applied is $\theta_i = \arccos \sqrt{f(i)}$. 
The following equation shows how each additional qubit rotation results in the doubling of probability regions in the quantum state, which represent the left and right halves of our PMF: $\sqrt{P(i)}|i\rangle\left|\theta_i\right\rangle|0\rangle \rightarrow \sqrt{P(i)}|i\rangle\left|\theta_i\right\rangle\left(\cos \theta_i|0\rangle+\sin \theta_i|1\rangle\right)$. 

\begin{figure} [ht]
\begin{center}
\begin{tabular}{c}
\includegraphics[scale=0.5]{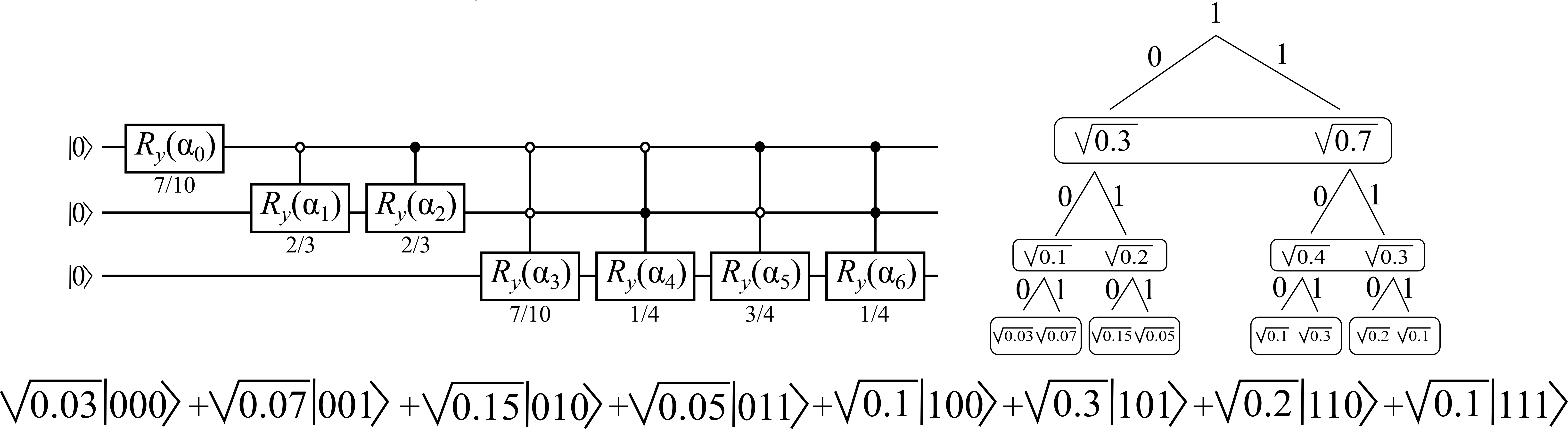}
\end{tabular}
\end{center}
\caption[example] 
{\label{fig:QRNG_example}
Example Distribution}
\end{figure}

The quantum states referenced above are generated using an amplitude circuit generation algorithm that computes rotations corresponding to the probability ratio between left and right subsections. For example, the first rotation controls the ratio between the entire left and right probability masses of the PMF; the next controlled rotation controls the ratio within the left side of the PMF.
From a classical algorithms perspective and visualized in Figure~\ref{fig:QRNG_example}, this method creates a binary tree where every layer of the tree has node values normalized to unity, and the values at the leaf nodes are the final ordered PMF probabilities. The ratio between the sum of all nodes on the left side and all on the right side determines the rotations at each parent node. Figure~\ref{fig:QRNG_example} illustrates the relationships between the amplitude encoding circuit and the corresponding amplitudes. This general amplitude encoding generation is created from multiple controlled-Pauli-$\mathbf{Y}$ rotation gates denoted as $CR_y$ with single qubit rotations about the $y$-axis denoted as $R_y(\alpha_i)$ with $\alpha_i$ denoting the angle of rotation. We will show how this algorithm is modified to work with PMFs in Sec.~\ref{sec:implementation_optimization_flex_qrng_circuits}

\subsection{QRNG Circuit Synthesis and Optimization}
Since the state preparation circuit that realizes the PMF is in the same form as a generalized state $\sum_x e^{i\phi} \sqrt{a_x}|x\rangle$, but without the complex phases, we can use a general quantum circuit to initialize a general state generating circuit that can then be optimized specifically for PMFs.
A baseline version of this procedure that we will use can be found in the \texttt{Qiskit.Initialize} function that implements the optimization methods in Ref.~\citenum{shende2005synthesis} and can generate a state generation circuit from specified probability amplitudes.
This circuit structure can synthesize the quantum circuit that results in amplitude encoding. Our Flex-QRNG allows us to enhance this initial circuit with significant additional optimizations.  Most notably, we adapt the gray-code optimization technique\cite{shende2005synthesis} for our purposes in producing the Flex-QRNG tool. 
The depth grows exponentially with respect to the number of qubits ($n$), meaning the circuit complexity is $O(2^n)$.

In addition, classically computing the rotations that comprise the circuit is a recursive process with complexity $O(2^n)$. 
So, there are other methods to address this limitation.
For example, a recent method in Ref.~\citenum{holmes2020entanglement} uses Monte Carlo methods, such as matrix product states (MPS), to reduce the classical calculation of rotations such that both the quantum circuit growth and classical computation time are linear in average time complexity. 
We will assume the computation impact of the classical preprocessing step is minimal because a QRNG will only need to be synthesized once per use case.

We briefly note that a different approach to synthesis and optimization of QRNGs employs variational algorithms.
Variational methods are useful for the noisy, intermediate-scale quantum (NISQ) era, given their relatively low depth and ability to adjust for hardware errors, given their development on noisy quantum channels.
Ref.~\citenum{zoufal2019quantum
} describes using variational methods to generate QRNGs.
However, this approach is only an approximation of the amplitude encoding method that we employ in Flex-QRNG. 
Furthermore, Ref.~\citenum{zoufal2019quantum
} does not analyze performance benefits with as many qubits, statistics, or distributions as are included in the remainder of this paper.
For example, Ref.~\citenum{zoufal2019quantum
} does not differentiate sampling error from hardware error when determining how both impact the shape of the distribution, nor does it include automation of a user-specified PMF to a specific QC hardware backend.
So, although the variational approach of Ref.~\citenum{zoufal2019quantum}
has a circuit gate-growth complexity of $O(\text{poly}(n))$, we will not further consider variational approaches in this work.
We instead show that the gate-count complexity of amplitude encoding---which offers an exactness not available from variational approaches---can be improved with optimizations that leverage knowledge of the distribution and can thus be mapped to multiple hardware backends that are further optimized by our Flex-QRNG tool.

\subsection{Differentiating Between States Representing Data and States Representing Distributions}
To differentiate between a state representing indexed data samples and a state representing a distribution, we can look to quantum machine learning (QML).
While both types of states can be stored in the amplitudes of a multi-qubit system using a similar circuit structure, they differ in how the general state representation is interpreted.
In QML, the state is known as a quantum probability distribution and is referred to as a quantum sample or qsample~\cite{schuld2021supervised}. 
In what is known as an amplitude-encoded circuit, each probability is interpreted as a data point, and the basis measured is just an index: $\sum_x \sqrt{a_x}|x\rangle$. Whereas in a qsample (quantum distribution) encoding circuit,
the output values align with a PMF: $\sum_x \sqrt{P(x)}|x\rangle$.
For a more generalized state initialization that is application agnostic, we can also include initializing phase values $\sum_x e^{i\phi} \sqrt{a_x}|x\rangle$ where $\lVert x \rVert_2=1$.
This equation shows that representing classical distributions in a quantum state can be accomplished using any general state synthesis algorithm, as we can ignore the complex phases.

There have been several past research papers that combine quantum computing and probabilistic learning models. 
Some methods combine QML with probabilistic, inference-based learning, such as Bayesian networks~\cite{low2014quantum}, hidden Markov, and Boltzmann machines\cite{amin2018quantum}.
These QML methods are examples of quantum algorithms that would benefit from having a specific QRNG compiler, such as that used as a basis for Flex-QRNG, to generate optimized subcircuits.

\section{Implementation and Optimization of Flex-QRNG Circuits}\label{sec:implementation_optimization_flex_qrng_circuits}

Before describing our specific QRNG-generation process, we briefly introduce the \textit{MustangQ} compiler, emphasizing the features that directly contribute to its usage as a Flex-QRNG automated design tool.
More complete and detailed descriptions of \textit{MustangQ} are provided in Refs.~\citenum{ST:17, ST:19a, ST:19b, SHT:22}.
\textit{MustangQ} is comprised of two portions: a ``front-end" that converts irreversible (and reversible) specifications into optimized technology-independent reversible circuits using a reversible gate set and a ``back-end" that maps the reversible circuit to a technology-dependent gate set based on a user-specified gate library, and then optimizes the technology-mapped circuit and generates an OpenQASM output file.
The portions of \textit{MustangQ} that specifically apply to its use as a Flex-QRNG generator related to the synthesis of discrete PMF are described in this paper. The initial structure that generates the amplitude encoded values is partially implemented in the \texttt{Qiskit.Initialize} function. The initialize function is related to the general synthesis of a quantum unitary gate. The amplitude encoding circuit in which the quantum states in superposition have a select non-equiprobable amplitudes -- amplitudes that in our research we choose to represent as PMF bins. We restrict the possible amplitudes of the initialize function to the real part of the amplitudes that allow us represent the real probabilities only. We have also implemented other experimental methods to assess the effect of sampling and bias errors when modeling common classical distributions on a quantum computer.

\begin{figure} [ht]
\begin{center}
\begin{tabular}{c}
\includegraphics[width=0.75\textwidth]{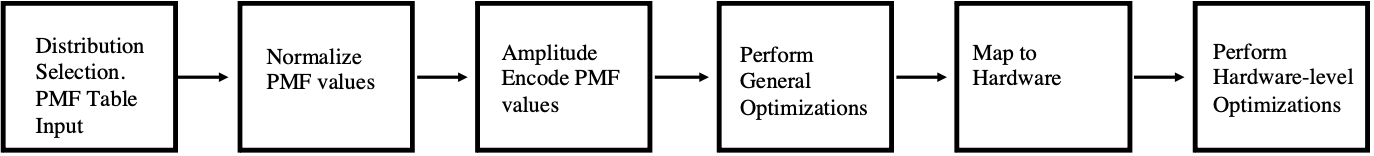}
\end{tabular}
\end{center}
\caption[example] 
{\label{fig:QRNG_process_flow} 
An overview of Flex-QRNG Generation.}
\end{figure}

The steps for generating a Flex-QRNG with \textit{MustangQ} are listed in Figure~\ref{fig:QRNG_process_flow}.
First is PMF selection and specification; the distributions can be either parametric---meaning described by a limited number of specific statistical parameters---or non-parametric or, as we will sometimes term them, arbitrary.
For example, the normal distribution is a parametric distribution because the variance and the mean can describe it. In contrast, some multimodal distributions can only be described with either complex parameters, a large number of parameters, or both.

The next step is to specify our PMF, which can be written in a tabular form containing floating point values.
Using our tool, we can represent both parametric and non-parametric distributions; even when it is difficult to characterize a distribution using statistical parameters, we can directly represent the distribution with a PMF table that contains bin heights.

To automate the creation of fixed-point tables containing the bin heights of parametric PMF distributions, we first use the \texttt{SciPy} library, which contains many common PMF functions.
The number of bins used depends on the number of qubits used: $N$ qubits correspond to $2^N$ PMF bins.
For non-parametric functions, we cannot use \texttt{SciPy} to automatically create a PMF, but the user can specify whatever PMF they would like by preparing a table analogous to that described above.

After defining PMF bins, we must interpret the bin heights as amplitudes, which requires normalizing the values such that the sum of the squares of the values equals one. To do this, we can divide by the $\mathcal{L}_2$ norm, which provides probability amplitudes $P(i)$, where $P(i) = \frac{x_i}{\sqrt{\sum_{j=1}^N x_j^2}}$.

After normalizing the probability amplitudes, the values are used in the amplitude encoding circuit to compute the conditional rotations for each subsection recursively.
These rotational parameters are then integrated into the circuit. This general amplitude encoding found in \texttt{Qiskit.Initialize} is not specific to a specific QC hardware. The usage of Flex-QRNG is needed to allow us to map to different varieties of backends. The process of synthesis involves using optimizations on the initial amplitude encoded circuit and then further mapping to specific QC hardware. This is followed by further backend hardware-specific optimizations. The Flex-QRNG compiler backend can easily be extended to target other platforms such as transmon, ion-trap, or photonic computers. This tool can be used to set up a comparison with different backend hardware since some backends will be faster and more efficient than others.

After this machine-dependent synthesis, the circuit can be mapped to specific hardware for simulation or execution.
(See Secs.~\ref{sec:experimental_setup} and ~\ref{sec:experimental_results} for discussion of such simulation.)

\begin{figure} [ht]
\begin{center}
\begin{tabular}{c}
\includegraphics[scale=0.5]{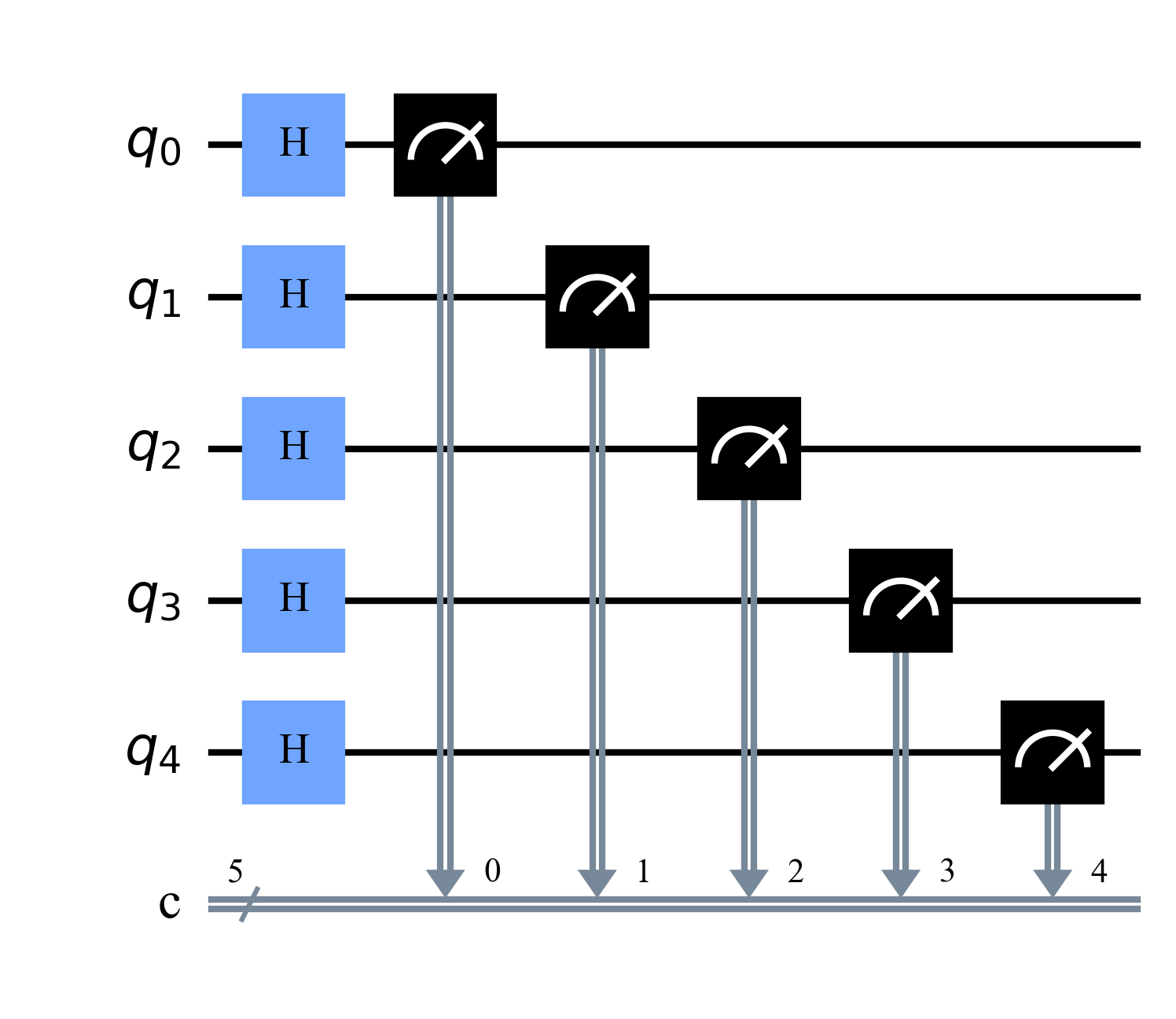}
\end{tabular}
\end{center}
\caption
{\label{fig:flex_QRNG_uniform_circ} 
A \textit{MustangQ}-generated Flex-QRNG for a 5-qubit uniform PMF.}
\end{figure}

Upon completing these steps, we have a Flex-QRNG circuit specified in a machine-independent language (OpenQASM) or as a hardware-dependent specification file.
The user needs only specify a distribution in the form of a PMF table and the number of qubits to generate a circuit, such as an example in Figure~\ref{fig:flex_QRNG_uniform_circ}, which is for the simplest option: a uniform PMF.

\subsection{Using FlexQRNG for the Discrete Uniform PMF}
To explain the above steps in more detail, we consider the Flex-QRNG-generation process for generating the very simple discrete uniform PMF as illustrated in Figure~\ref{fig:flex_QRNG_uniform_circ}.
A discrete, uniform PMF generates random samples that are equiprobable.
A uniform PMF is the simplest case in terms of specifying the PMF since it is a constant function: $P_{uniform}(X)=p$, where $X$ is a discrete RV whose variates are all possible values of the $N$ qubits used in the Flex-QRNG.
Likewise, $p$ is the probability that any possible $N$-qubit value will result when the Flex-QRNG is executed.
So, for the case in which $N=5$, the set of variates for $X$ comprises all $2^5=32$ different bitstrings: ${0\times00,0\times01,...,0\times FE,0\times FF}$. 

Ordinarily, the next steps would be defining the distribution in tabular form, normalizing the probability amplitudes, and recursively computing rotational parameters.
However, for the case of the uniform distribution, the tabular form is trivial (all probability amplitudes are the same), the probability amplitudes are already normalized, and no rotations are required.
Therefore, we move straight to circuit synthesis.
After initializing the qubits to $\ket{0}$, the next portion of the circuit encodes the desired PMF into the probability amplitudes of the qubits.
One way to achieve the uniform PMF case is to apply a bank of five single-qubit Hadamard gates to each initialized qubit, which evolves the qubits into an amplitude-encoded uniform PMF as given in Eq.~\eqref{eq:hadamards_applied}.
\begin{equation}
\label{eq:hadamards_applied}
H^{\otimes 5}\ket{ q_4 q_3 q_2 q_1 q_0} = \frac{\ket{00000} + \ket{11111}}{\sqrt{2}^5}
\end{equation}
Then, a sample of the RV variate is obtained by simultaneously measuring all the encoded qubits.
The measurement causes each qubit to collapse into a conventional bit of information in the set $\mathbb{B}=\{0,1\}$, with a probability equal to the squared amplitude of the encoded probability amplitudes of each qubit.
It is worth noting that the phase of each qubit is irrelevant, which offers a degree of freedom that the compiler can exploit to enable circuit optimizations.
The bitstring resulting from measurement can be interpreted in several ways.
For example, if the user desires a serial random bit stream, the five measured qubit values can be used individually.
Alternatively, suppose a higher-radix variate is desired. In that case, the resulting measurement can be interpreted as two (or three) digits in the binary number system, either from $\{0, 1, 2, 3, 4\}_{10}$ or $\{0, 1, 2, 3\}_{10}$, depending upon how many digits are used per number.
Similarly and more generally, a substring of the measured qubits can be interpreted as the binary representation of a higher-radix digit as long as the length of the substring is an integer factor.
Clearly, any fixed-point value may likewise be interpreted, or the bit strings may be interpreted as signed values.

\subsection{Using FlexQRNG for Discrete Non-Uniform PMFs}
Synthesizing a Flex-QRNG circuit for a discrete uniform distribution can be automated using our synthesis tool.  In this simple case of an equiprobable PMF the user would need to only specify the word size of the uniform distribution. The compiler would map the Hadamard gates to specified qubits depending on the hardware backend.
For more complex PMFs, using the automated tool is an even more straightforward---and less error-prone---approach than manually designing the state generation circuit.
One example of a more-complex distribution is the binomial PMF, which is required for some post-quantum cryptography algorithms~\cite{VL:12,LD:13}.
We consider generating the binomial PMF using a Flex-QRNG here.

The binomial PMF generation circuit should evolve to have probability amplitudes that are proportional to the positive square root of the binomial coefficients.
The binomial coefficients depend on the total number of qubits, $N$, chosen for the Flex-QRNG and are the scalar coefficients of the expanded polynomial for $(x+y)^N$.
For example, in the case of $N=5$, these coefficients are the ordered set, which are then normalized as probabilities of a PMF. The example shown in the lists Figure \ref{fig:list_coefficients}.

\begin{figure} [ht]
\label{fig:list_coefficients}
\begin{align*}
\text{binomial}&=\numberlist[0.75\linewidth]{
1,31,465,4495,31465,169911,736281,2629575,7888725,20160075,44352165,84672315,141120525,206253075,265182525,300540195,300540195,265182525,206253075,141120525,84672315,44352165,20160075,7888725,2629575,736281,169911,31465,4495,465,31,1
}\\
\text{binomial normalized} &=\numberlist[0.75\linewidth]{
1.4658E-09,4.54397E-08,6.81595E-07,6.58875E-06,4.61213E-05,0.000249055,0.001079238,0.003854421,0.011563263,0.029550561,0.065011233,0.124112355,0.206853925,0.302324967,0.388703529,0.440530666,0.440530666,0.388703529,0.302324967,0.206853925,0.124112355,0.065011233,0.029550561,0.011563263,0.003854421,0.001079238,0.000249055,4.61213E-05,6.58875E-06,6.81595E-07,4.54397E-08,1.4658E-09
}
\end{align*}
\end{figure}

\begin{figure} [ht]
\begin{center}
\begin{tabular}{c}
\includegraphics[scale=0.3]{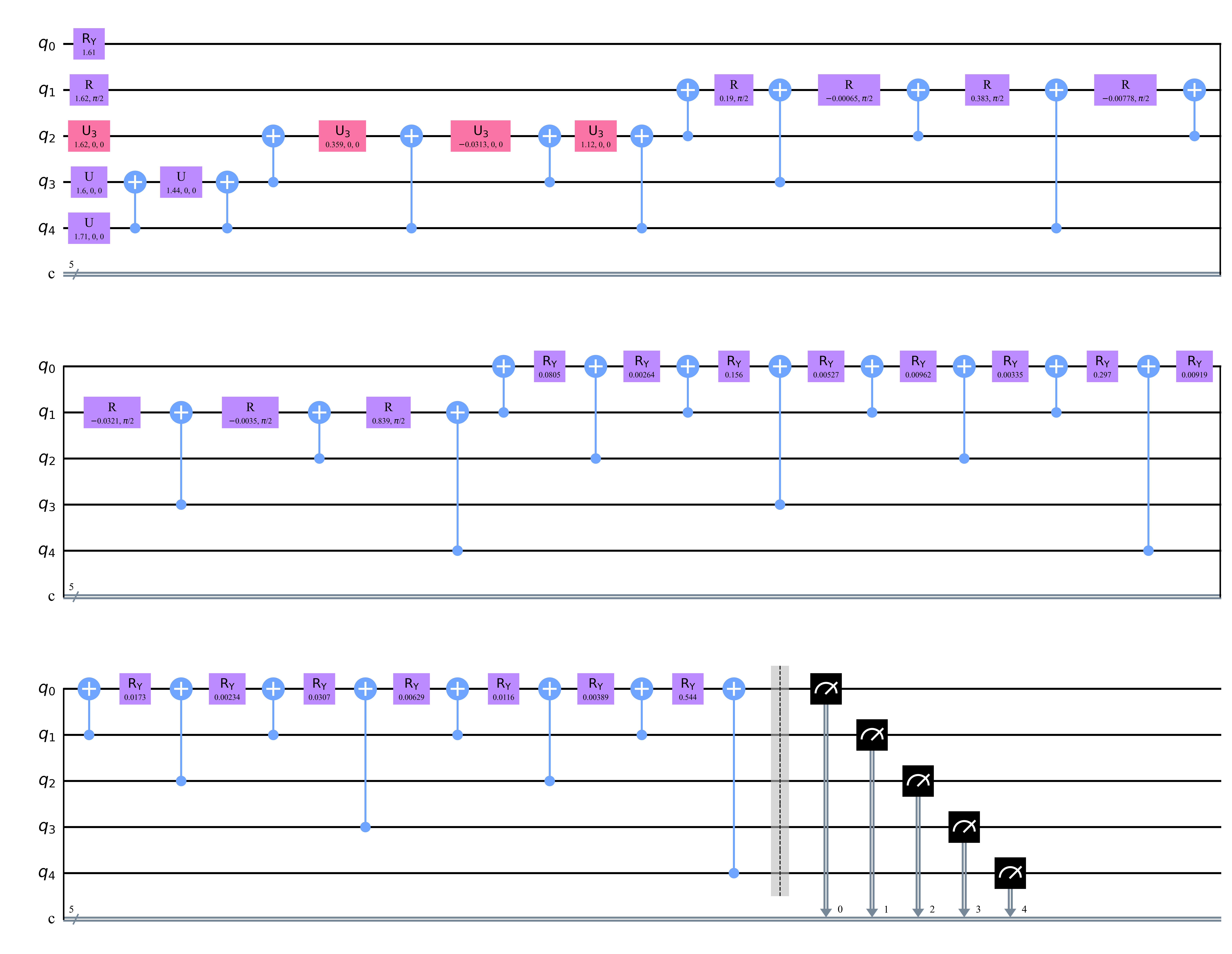}
\end{tabular}
\end{center}
\caption[example] 
{\label{fig:flex_QRNG_binom_circ} 
Flex-QRNG for 5-qubit Binomial PMF.}
\end{figure}

Even for such a well-defined PMF, manually designing and optimizing a state generation circuit is non-trivial, illustrating the utility of \textit{MustangQ} configured to function as FlexQRNG.
We chose $N=5$ qubits and the Clifford+\textbf{T} single-qubit gate set augmented with the controlled Pauli-\textbf{X} gates to form a universal set.
The resulting \textit{MustangQ}-generated Flex-QRNG circuit is shown in Figure~\ref{fig:flex_QRNG_binom_circ}.

\subsection{Using FlexQRNG for Discrete Non-Parametric PMFs}

Generating samples from an arbitrary, non-parametric distribution is a challenge in classical random number generation. A nonparametric distribution is defined as a data distribution that cannot be represented through a standard, parameterized probability distribution. Classical random number generation primarily focuses on generating uniform random numbers and occasionally more complex parametric distributions such as normal distributions. There are some ways to create arbitrary distributions classically, but these are inefficient. The rejection sampling algorithm, commonly used in statistics, is a simple way of sampling from any arbitrary distribution using samples from a uniform generator. The technique involves throwing out or rejecting the samples that fall out of a predefined range of values given a function that defines the shape of the distribution.

The excess computation that is wasted from rejection sampling results in a lack of efficient arbitrary PMF generation on classical computers. For this reason, using the Flex-QRNG to create non-parametric PMFs would be a key contribution to the area of random number generation. The rapid generation of values from a non-parametric distribution would vary depending on backend hardware used. Nevertheless, the lack of wasted measurements and ability to program the distribution in the quantum algorithm directly gives it an efficiency advantage. When the generation efficiency is combined with the fact that the numbers generated are true random numbers, we obtain a uniquely flexible tool that can be used in many scenarios.

\section{Experimental Setup}\label{sec:experimental_setup}
To illustrate the automated generation of non-trivial Flex-QRNGs, we use \textit{MustangQ} to synthesize five PMF distributions: uniform, binomial, triangle, bimodal, and arbitrary.
As described in Sec.~\ref{sec:implementation_optimization_flex_qrng_circuits}, we then use the algorithm found in \texttt{Qiskit.Initialize} restricted to real amplitudes to represent the initial amplitude encoded dataset. This initial circuit is made up of high-level multiple-controlled $R_y$ gates with parameters of rotation. We then use \textit{MustangQ} to represent this circuit in a way that can be further optimized and mapped to various technologically-dependent gate sets including those that support various IBM QC.

Once we have a synthesized the QRNG circuit, we simulate it using both an ideal simulator and a noisy simulator designed to replicate the particular noise signature of specific IBM-Q hardware~\cite{NoiseModel:23}.
Finally, we analyze the quality of the random numbers generated by the Flex-QRNG circuits when simulated without and with noise.
In this section, we provide a brief introduction to the statistics used and a pre-experiment step that ensures inadequate sampling does not adversely affect our results.  We note that we use the QC simulators since we did not have access to actual QC with a sufficient number of qubits.  We did verify that our QRNG state generation circuits successfully execute on actual QC hardware for cases wherein a very small number of qubits we required.

\subsection{Experimental Result Analysis}\label{sec:experimental_result_analysis}
The Kullback–Leibler divergence (KL divergence)---shown in Eq.~\eqref{eq:kl_divergence} for discrete probability distributions --- measures how much an empirical distribution, $Q$, `diverges' from a theoretical distribution, $P$.
We can thus apply KL divergence to assess the similarity of the distributions from a Flex-QRNG circuit and the PMF that circuit is synthesized to represent.

\begin{equation}
\label{eq:kl_divergence}
\operatorname{KL}(P \| Q)= \sum_i P(i) \ln\left(\frac{P(i)}{Q(i)}\right)
\end{equation}

To understand what causes the differences in the distributions that divergence identifies, we need to briefly consider quantum error and how it affects different simulators.
We consider two broad classes of quantum error that we refer to as ``sampling error'' and ``hardware error.''
Sampling error is a consequence of quantum computing's probabilistic nature; even theoretically-perfect quantum hardware is susceptible to sampling error.
Conversely, it is generally thought that most hardware errors will eventually be avoidable because this type of error is due to interference or decoherence that results from the difficulties of building quantum computers~\cite{NC:00}.
Ideal simulators thus include sampling error but no hardware error, while noisy simulators include both sampling and hardware error.
(It is worth clarifying that by ``ideal simulator," we are \textit{not} referring to a ``statevector simulator," which does not provide output in the form of measurements but instead provides a vector representation of the circuit state for sufficiently small circuits.)

The goal of our experiments is twofold: first, to illustrate that the Flex-QRNG circuits correctly generate the PMF distributions they are designed for, and second, to assess the adverse effects of hardware noise on Flex-QRNG output.
These analyses require separating the effects of sampling and hardware noise as much as possible.
Additionally, the first task of assessing whether the Flex-QRNG circuit accurately represents the desired PMF requires using a sufficient number of shots; otherwise, the circuit might be mischaracterized as poorly representing the PMF when the real issue is simply one of inadequate sampling.
Thus, prior to simulating the Flex-QRNG circuits, we determine the number of shots that reduces sampling error by performing a ``G-test,'' which is a statistical test derived from KL divergence.

To compute this shot count, we need a quantifiable test statistic for measuring the significance of the number of shots on the distribution.
We use the Goodness-of-fit-test, or G-test, between ideal simulation results and the PMF the circuit should represent.
KL divergence is proportional to the G-test, so we can use KL divergence to determine shot count by interpreting the value as a G-test statistic~\cite{mcdonald2009handbook}.
Unlike most statistical experiments that want to differentiate two distributions, we aim to ensure the distributions are the same. Thus, our goal is to minimize the G-test result, known as the G-statistic.
This is because the null hypothesis states that the distributions are the same, so minimizing the G-stat to zero means that we \textit{cannot} reject the null hypothesis that the two distributions are the same.

In other words, a small G-stat indicates that the distribution of random numbers from the Flex-QRNG circuit is not statistically different from the PMF from which it was synthesized. It is useful to note that the G distribution is approximately the same as the Chi-square distribution, so for all our experiments, we can use our G-statistic to obtain a readily interpretable P-value as the probability that our sampled distribution is the same as the theoretical distribution.

As shown in Eq.~\eqref{eq:G_stat}, the G-stat can be written as the KL divergence times twice the number of samples~\cite{mcdonald2009handbook}.
Our experiment allows us to interpret the p-values from the G-stat as an indicator of the probability that the samples from our empirical distribution are from the theoretical distribution.
Specifically, decreasing the G-test value to zero is equivalent to maximizing the p-value to 1.
(A p-value of 1 indicates that we should reject the hypothesis that there is a difference between the distribution given by the circuit running on the ideal simulator and the PMF we seek to model.)

\begin{equation}
\label{eq:G_stat}
\begin{split}
G &= 2 \sum_i O_i \ln\left(\frac{O_i}{E_i}\right) \\
&= 2 \sum_i n P(i) \ln\left(\frac{n P(i)}{n Q(i)}\right) \\
&= 2n \sum_i P(i) \ln\left(\frac{P(i)}{Q(i)}\right) \\
&= 2n \times KL(P \| Q)
\end{split}
\end{equation}

To use the G-stat to determine shot count, we collect preliminary data using the ideal simulator.
Specifically, we simulate the Flex-QRNG with increasingly many shots until our sampling error becomes negligible according to a G-test.
We then increase this number of shots by an additional 50\% to avoid underestimation.
For all experiments, we use this number of shots when assessing how similar the Flex-QRNG results are to the PMF distribution.

Table~\ref{table:shot_selection} shows that distributions which have more variation are also more difficult to sample from and, thus, require more shots.
Consequently, distributions with large peaks and valleys will require more samples to get more resolution from the distribution.
To generate the results in Table~\ref{table:shot_selection}, we simply increased the number of shots until the G-stat fell below $1E-3$, which corresponds to a p-value close to one for all distributions.
Using this computed number of shots (with some additional shots to further reduce the margin of error), we can estimate the divergence of the noisy simulation from the ideal distribution without having sampling error mixed in. 

\begin{table}[ht]
\begin{center}
\begin{tabular}{|c | c c c c || c|} 
 \hline
 & KL-divergence & JS  & G-stat & G-stat P-Value & Shots(x1.5)\\  
 \hline
 Uniform & 0.0006 & 0.0121 & 0.0012 & 0.9723 & 34500 \\
\hline
Binomial & inf & 0.0129 & 0.0013 & 0.9712 &  21000 \\ 
\hline
Triangle & 0.0005 & 0.0113 & 0.001 & 0.9747 & 34500 \\ 
\hline
Bimodal Non-parametric & 0.0005 & 0.0111 & 0.001 & 0.9747 & 36000 \\ 
\hline
Arbitrary Non-parametric & 0.0004 & 0.0101 & 0.0008 & 0.9774 &  42000 \\ 
\hline
\end{tabular}
\caption{\label{table:shot_selection}
Selecting a number of shots for each distribution to reduce sampling error when using the ideal simulator.}
\end{center}
\end{table}

Before describing the results, it is worth making two additional notes about the statistics in our analysis.
First, a Chi-square test provides approximately the same information as a G-test.
However, G-tests have the benefit of additive results~\cite{mcdonald2009handbook}, meaning we can easily combine multiple experiments that use different numbers of samples or even subtract sources of sampling error.
Additionally, the G-test is derived from the KL-divergence, which is an intuitive way to compare our Flex-QRNG results to the original PMF distributions.
However, because the Chi-square test is more common, we compute those statistics as well.

Second, despite the intuitive benefit of KL divergence, it is not a true distance metric because it is not symmetric.
This means that if the theoretical distribution and the empirical distribution are swapped in Eq.~\eqref{eq:kl_divergence}, the value of the expression changes.
Furthermore, the KL divergence can be infinite in cases where the empirical distribution does not overlap with the theoretical distribution. This can be shown in the table \ref{table:shot_selection} that the binomial distribution has a KL-divergence of infinity while the other measures are straightforward to interpret.
Therefore, we also consider the Jensen-Shannon divergence (JS divergence) in Eq.~\eqref{eq:js_divergence}, which eliminates the aforementioned problems by taking the average of both combinations of KL divergence with the theoretical and empirical distributions.

\begin{equation}
\label{eq:js_divergence}
\operatorname{JSD}(P \| Q)=\frac{1}{2} KL(P \| M)+\frac{1}{2} KL(Q \| M) \text{ where } M=\frac{1}{2}(P+Q)    
\end{equation}

\section{Experimental Results}\label{sec:experimental_results}
This section presents results to both illustrate the functionality of the Flex-QRNG circuits and to optimize the circuits for better performance.
We first consider whether the Flex-QRNG, absent any hardware noise, accurately represents the PMF that it should.
Next, we consider the Flex-QRNG's distributions in the presence of simulated hardware noise.
Then, we introduce circuit optimizations to improve Flex-QRNG performance.

\subsection{Ideal Simulation Versus Desired PMF}
\begin{figure}[ht]
    \begin{center}
    \begin{tabular}{c}
    \includegraphics[width=\textwidth]{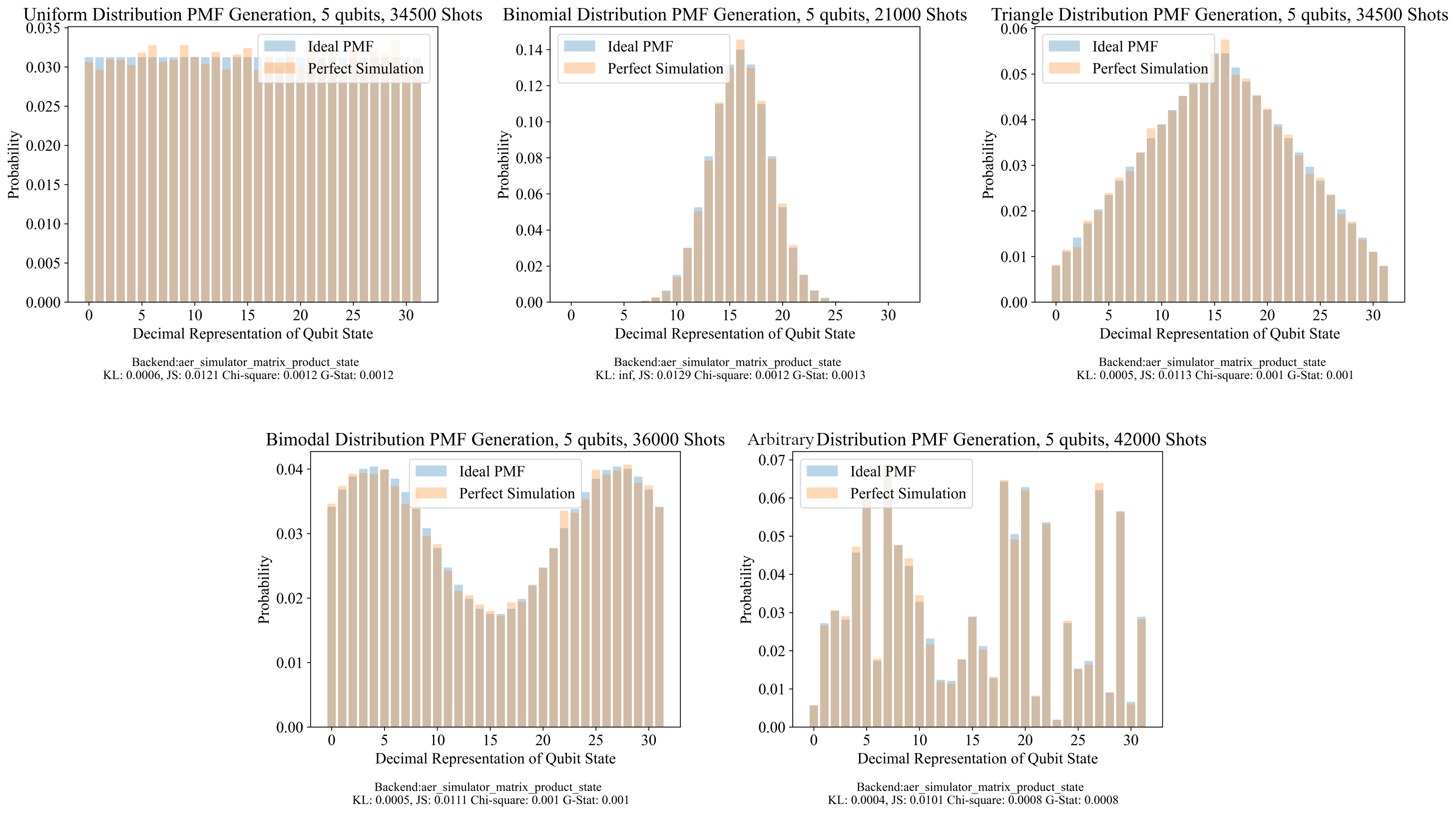}
    \end{tabular}
    \end{center}
    \caption{\label{fig:noiseless_simulations}
    Generated PMFs from Flex-QRNG circuits simulated using an ideal simulator.  The blue histogram illustrates the PMF input to the Flex-QRNG generation process, while the orange histogram illustrates the distribution measured when simulating the Flex-QRNG circuit.}
\end{figure}

For the PMFs generated in this section, we applied the methodology described above to determine the number of shots to use.
We synthesized Flex-QRNGs for both parametric distributions (uniform, triangle, and binomial PMFs) and non-parametric PMFs (bimodal and arbitrary distributions) illustrated in Figure~\ref{fig:noiseless_simulations}).
For each circuit, we chose to use five qubits for two reasons.
First, simulating five qubits is computationally manageable; simulating more than five qubits requires significantly more time, including because the hardware-routing task becomes more complex.
Second, nearly all of IBM-Q's hardware has five qubits, meaning these results can---in future work---be compared to results from running the Flex-QRNGs on actual quantum hardware.
Upon synthesizing a circuit for each distribution, we import the OpenQASM specification~\cite{CB+:17,CB+:20} into the Qiskit environment and evaluate each Flex-QRNG by executing an acceptable number of shots computed using the approach in Sec.~\ref{sec:experimental_setup}.

These four examples illustrate the utility of our approach for generating the Flex-QRNG quantum circuits automatically.
\textit{MustangQ} offers flexibility, given that any number of qubits can be specified; the interpretation of the variates can be assumed as values from various digit sets, including fixed-point and signed deals, and any arbitrary PMF can be specified.
The automated synthesis tool also lets users specify any target quantum gate set.
It thus supports different quantum hardware or can be used as a design tool for an application-specific device, such as an integrated circuit based upon superconducting semiconductor qubits.

\subsection{Noisy Simulation Versus Desired PMF}
We now assess how well Flex-QRNG circuits generate the desired PMFs in the presence of simulated hardware noise.
IBM provides simulators that approximate the noise signatures of actual hardware, and for our experiments, we chose the IBM-Q Washington machine, which is amongst the most sophisticated of IBM's suite of quantum devices.
We used the default five qubits as determined by the IBM qiskit transpiler. The transpiler uses machine information such as qubit connectivity and qubit error information.

\begin{figure}[ht]
    \begin{center}
    \begin{tabular}{c}
    \includegraphics[width=\textwidth]{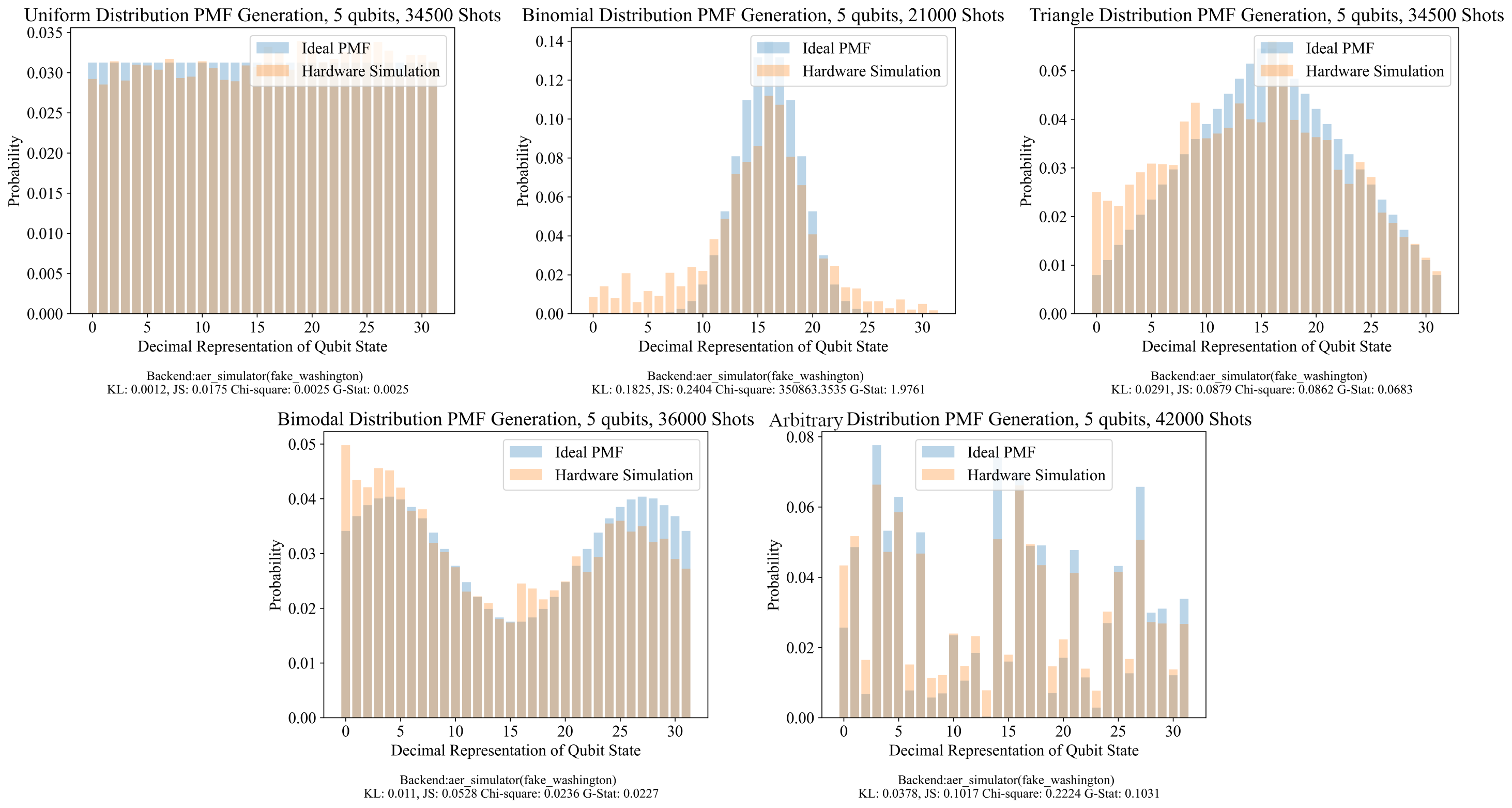}
    \end{tabular}
    \end{center}
    \caption{\label{fig:noisy_simulation_washington}
Generated PMFs from Flex-QRNG circuits simulated using a noisy simulator of the IBM-Q Washington.  The blue histogram illustrates the PMF input to the Flex-QRNG generation process, while the orange histogram illustrates the distribution measured when simulating the Flex-QRNG circuit.}
\end{figure}

Figure~\ref{fig:noisy_simulation_washington} contains the histograms and computed statistics for the noisy simulation.
The $x$-axis specifies the number of qubit states, while the $y$-axis specifies occurrence frequency.
Even without considering the numerics, the plots illustrate that the distributions generated by the Flex-QRNG in noisy simulations have significantly greater errors than those generated when using an ideal simulation of the same distribution.

We can now discuss and compare the generated distributions using the KL divergence, JS divergence, Chi-Square, and G-stat statistics.
First, considering the uniform distribution in Figure~\ref{fig:noisy_simulation_washington}, we can see the bias due to hardware error.
In fact, we can see the error visually as the right half is higher than the left half.
For the uniform distribution, the G-stat results in a p-value around 0.96, which means that it has high goodness of fit and that there is a 96 percent probability that the empirical dataset is sampled from the original.
The low error of this distribution circuit is due to the low depth and, thus, the lower probability of decoherence.

For the binomial distribution, the p-value from the G-stat is around 0.16.
The reason for this major drop can be seen from the penalty that occurs when there is an event that does not exist in the theoretical PMF from which a Flex-QRNG circuit is derived.
In this case, we still find that the JS divergence, KL, and Chi-square are more than three times higher for the binomial distribution than for any other distribution.

For the triangle distribution, the p-value from the G-stat is around 0.794.
One possible reason for this relatively worse result is that the triangle distribution has lower probabilities on the tail ends of the distribution. 
These low probabilities may be hard to represent, as they must be synthesized using very small rotations on a quantum computer.

For the next distribution---the bimodal distribution--- the p-value from the G-test is approximately 0.880.
This non-parametric distribution has a significantly higher p-value than all distributions other than uniform, which means that the probability that the output distribution is sampling from the right theoretical distribution is more likely.
For the next non-parametric distribution, the arbitrary distribution, the p-value from the G-stat is approximately 0.75, indicating about as much error as for the triangular distribution.

Table~\ref{table:synthesis_comparison_statistics} provides metrics regarding the synthesized Flex-QRNG circuits after decomposing to the native IBM QC $\mathbf{U3}$ and $\mathbf{CX}$ gates. Metrics include the number of gates used and quantum depth.
Quantum depth can be defined in a variety of ways; here, we report the critical path length computed by Qiskit~\cite{QIS:21}.
These results illustrate that distributions without small values are more accurate than flatter ones. The primary example is that the noisy binomial and triangle distributions have the lowest p-values, whereas the bimodal and arbitrary distributions have higher p-values.
This is likely a consequence of the reduced precision of rotations when moving from testing on ideal simulations to hardware simulations.
This reduction in precision is due to the noise in quantum hardware rotations caused by a combination of gate errors and decoherence.
As described in \cite{Sinha2022}, reduced precision may result in underflow for lower probabilities, and as a result, distributions have more errors on lower probability bins than others. The previous method \cite{Sinha2022} discusses methods to improve the underflow that occurs for other data representations but does not extend to the quantum random distributions discussed in this work.

If we order the distributions by any of the metric performances, the order will all be the same; in other words, all metrics increase and decrease together across different distribution types.

\begin{table}[ht]
\begin{center}
\begin{tabular}{|c | c c c c || c  c|} 
\hline
& KL-divergence & JS  & G-stat & G-stat P-Value &  Gates &  Depth \\  
\hline
Uniform & 0.0012 & 0.0175 & 0.0025 & 0.9601224 & 5 & 1 \\
\hline
Binomial & 0.1825 & 0.2404 & 1.9761 & 0.1598019 & 61 & 57 \\ 
\hline
Triangle & 0.0291 & 0.0879 & 0.0683 & 0.7938283 & 61 & 57 \\ 
\hline
Bimodal Non-Parametric & 0.011 & 0.0528 & 0.0227 & 0.8802398 & 61 & 57 \\ 
\hline
Arbitrary Non-Parametric & 0.0378 & 0.1017 & 0.1031 & 0.7481408 & 61 & 57 \\ 
\hline
\end{tabular}
\caption{Statistical comparison of Flex-QRNG circuits and resulting distributions for five different PMFs when simulated on IBM-Q's Washington device.}
\label{table:synthesis_comparison_statistics}
\end{center}
\end{table}

\section{Conclusion}
This work introduces an automated design tool for generating flexible TRNGs suitable for execution on a general-purpose quantum computer.
Quantum computers are growing, with a 1000-qubit machine predicted by the end of this year~\cite{choi2022}.
As larger devices appear, our approach to straightforwardly generating flexible QRNGs that support arbitrary PMFs will allow for generating random numbers from a much larger range. 
If enough qubits are available, a discrete PMF could represent a sampled binomial distribution, that meets the NIST requirements.

Future work should investigate using the Flex-QRNG as a primitive in new cryptographic applications.
For example, Alice and Bob could use a shared one-time pad that defines the parameters of a Flex-QRNG---the number of qubits and the PMF---and the resulting Flex-QRNG could provide the basis of an encryption method.
Additionally, future work should explore the benefits of using quantum distributions and possible parallelized processing and probabilistic calculations on these distributions for statistical analysis.

The Flex-QRNG tool could be used to generate random numbers on multiple hardware technologies and compare efficiency of generation across multiple hardware providers. There are also other optimizations that can be developed specifically for PMFs that take advantage of the symmetric nature of parametric distributions.

\bibliography{report}
\bibliographystyle{spiebib}

\end{document}